# Saliency Inspired Quality Assessment of Stereoscopic 3D Video

Amin Banitalebi-Dehkordi and Panos Nasiopoulos
University of British Columbia, Vancouver, BC, Canada V6T 1Z4
Email: {dehkordi, panos}@ece.ubc.ca

*Abstract* —To study the visual attentional behavior of Human Visual System (HVS) on 3D content, eye tracking experiments are performed and Visual Attention Models (VAMs) are designed. One of the main applications of these VAMs is in quality assessment of 3D video. The usage of 2D VAMs in designing 2D quality metrics is already well explored. This paper investigates the added value of incorporating 3D VAMs into Full-Reference (FR) and No-Reference (NR) quality assessment metrics for stereoscopic 3D video. To this end, state-of-the-art 3D VAMs are integrated to quality assessment pipeline of various existing FR and NR stereoscopic video quality metrics. Performance evaluations using a large scale database of stereoscopic videos with various types of distortions demonstrated that using saliency maps generally improves the performance of the quality assessment task for stereoscopic video. However, depending on the type of distortion, utilized metric, and VAM, the amount of improvement will change.[1]

## I. INTRODUCTION

3D video technologies have entered the consumer market in the past several years. These technologies have not only affected specialized target areas such as entertainment, education, and medical imaging, but have also changed the quality of the viewing experience for the average consumer by bringing life-like 3D video into gaming, theater, mobile phones, and television [57], [62-63], [69-76], [80-84]. Therefore, it is crucial in 3D video content creation and delivery to design for providing the end-users with the highest possible quality of experience. The ultimate way to assess the delivered 3D quality is through subjective experiments, which are time consuming and expensive. An alternative solution is to develop objective quality metrics, in attempt to model the Human Visual System (HVS) to measure the perceptual quality. Considering that quality assessment metrics in 3D systems try to model the human visual system, it would naturally make sense that 3D quality metrics take into consideration visual attention models by considering the likelihood of image/video regions being looked at by the viewers.

Visual Attention Models (VAMs) imitate the HVS in predicting eye gazed points by estimating the likelihood of each region of an image to draw the attention. VAMs therefore have a diverse range of application in image and video compression [1-3], object recognition and detection [4-7], visual search [8], retargeting [9-10], retrieval [11-12], quality assessment [13], image matching [14], segmentation [15-16], and figure-ground separation [17]. Of these applications, one of particular importance is to use VAMs in the design of quality metrics where they guide the quality assessment techniques towards the most salient regions of an image or video. Then, quality metrics treat visible distortions in the salient regions differently than the ones existing in the non-salient regions [13,18-30].

Designing VAMs for 2D video goes back to over two decades ago and their applications are well explored [1-28]. Saliency measurement is already being integrated into various Full-Reference (FR: when a reference video signal is available for comparison) and No-Reference (NR: when no reference point is available) 2D video quality metrics. However, since 3D video technologies have entered the consumer market only in the past few years, the existing literature on 3D VAMs and their applications is yet to be as complete as the 2D case. Consequently, the application of 3D VAMs in designing 3D video quality metrics needs to be addressed. This makes more sense by considering the fact that quality measurement techniques usually calculate the amount of local visible distortions, similarities, or image statistics and perform pooling to generate the final metric value. Saliency maps resulted from VAMs can be integrated to the quality assessment pipeline in the pooling stage by emphasizing on the most salient regions in the content.

In order to integrate the saliency prediction into the full-reference stereoscopic quality assessment task, Zhang et al. proposed to use their 3D saliency map as a weighting factor for the Structural Similarity (*SSIM*) index [31] in Depth Image Based Rendering applications [32]. In their method, 3D saliency is modeled as an average of 2D image saliency and 2D saliency map resulted from depth map. In another work by Chu et al., saliency is extracted using 2D VAM of [33] and used as weighting factor in their quality metric design [34]. Jiang et al. [35] used the 2D spectral residual VAM [36] along with a method of foreground and background depth maps for saliency prediction on stereoscopic images. They used a hard-threshold value for the depth map to split it into foreground and background depth and combined these two maps with the result of 2D saliency map of [36]. A major drawback of the mentioned approaches is that 2D VAMs are used for saliency measurement of 3D content, while experiments have shown that 2D VAMs fail to accurately predict 3D human visual saliency [37-38]. In addition to using 2D VAMs, the methods mentioned above do not take into consideration the temporal aspects of the video as they are solely based on single image quality assessment.

[1] This work was partly supported by Natural Sciences and Engineering Research Council of Canada (NSERC) under Grant STPGP 447339-13 and Institute for Computing Information and Cognitive Systems (ICICS) at UBC.
   Authors are with the Electrical Engineering Department and ICICS at UBC, Vancouver, BC, Canada ({dehkordi, panos}@ece.ubc.ca).

In the case of no-reference stereoscopic video quality assessment, Gu et al. proposed to add a saliency term in the formulation of their sharpness metric [29]. This term is defined as linear correlation between the 2D saliency maps of the two views. In another work by Ryu and Sohn [30], a no-reference quality assessment method for stereoscopic images is proposed by modeling the binocular quality perception in the context of blurriness and blockiness [30]. In this method, 2D VAM of *GBVS* (Graph Based Visual Saliency) [33] is used for saliency evaluations. The NR metrics proposed by Gu et al. and Ryu and Sohn are both using 2D visual attention models to predict saliency for stereoscopic content. As mentioned, it is proven that 2D VAMs lack accuracy in saliency prediction for 3D data. Moreover, the two NR metrics are designed for quality assessment of stereoscopic images and do not consider the temporal aspects of the quality evaluation.

As explained, there is still a lack of saliency based NR and FR quality metrics specifically for *stereoscopic videos*. The existing methods are mostly based on using *2D image* saliency maps and do not take into consideration the temporal aspects of the video saliency and quality. The primary question that this study seeks to answer is whether saliency predictions from 3D Video VAMs can be used to improve the performance of full-reference and no-reference quality assessment of stereoscopic 3D video or not. Moreover, we are interested to know that in case of any performance improvements, how much gain can be achieved for each metric or VAM? Are these potential improvements dependent on the type of distortion they assess? To answer these questions, in this paper, we propose to use 3D VAMs for quality evaluation of 3D video, for both NR and FR cases. To this end, we integrate our previously designed Learning Based Visual Saliency (*LBVS3D*) prediction model for stereoscopic 3D video (along with several other VAMs) to various state-of-the-art FR and NR 3D video quality metrics. We evaluate the added value of incorporating 3D VAMs in FR and NR quality assessment of stereoscopic video using a large scale database of stereoscopic videos and 88 participants. The main contributions of this paper can be summarized as follows:
- Integrating saliency prediction maps from 5 different 3D VAMs to 13 full-reference and 12 no-reference existing 3D quality metrics. Depending on the design of each metric, the saliency integration is performed either in spatial or frequency domain (at potentially different pyramid scale levels). This basically means that the formulation of each metric is uniquely updated so that it treats salient regions of videos differently than non-salient areas.
- Extensive performance evaluations to identify if there are any improvements when saliency information is incorporated in quality assessment of stereo video:
    o This is done by comparing quality assessment performance of 3D metrics with and without saliency integration
    o Validation through a large scale stereoscopic 3D video dataset: 16 reference videos, each with 7 different kinds of distortions (at 13 levels), resulting in 208 stereo videos
    o Subjective evaluation using 88 subjects

The rest of this paper is organized as follows: Section II describes our saliency integration methodology, Section III contains details regarding our experiments, Section IV provides the results, and Section V concludes the paper.

## II. METHODOLOGY

Video quality metrics perform the quality assessment task by measuring similarities (when a reference is available) or distortion densities (when no reference is available) for partitions of the video and then combining the local partition measurements to an overall quality index in a process known as pooling. Quality pooling can be done spatially (for image quality assessment), or temporally (for video quality assessment). It is therefore possible, for these types of metrics, to incorporate saliency prediction results in the pooling stage of quality assessment pipeline. Fig. 1 shows the proposed saliency integration scenario. The rest of this section elaborates on the utilized 3D VAMs, various FR and NR quality metrics used in our experiments, and how to integrate saliency detection results in each quality metric.

*A. Learning Based Visual Saliency for 3D (LBVS3D)*

In our previous work [38], we designed a Learning Based Visual Saliency (*LBVS3D*) prediction model of attention for stereoscopic video. This model takes into consideration both low-level stimulus driven saliency features such as depth, motion, brightness, texture, and color, as well as high-level context dependent attributes such as presence of humans, text, vehicles, animals, and horizon. In addition, the effects of object size, compactness, sparsity, frame rate of stereoscopic video [39,40], and 3D visual discomfort are taken into account. Our eye tracking experiments showed that *LBVS3D* has close correlation with the human fixation data [38,41]. A block-diagram of *LBVS3D* (from [38]) is depicted in Fig. 2. It is observed from Fig. 2 that low-level and high-level features are both integrated within a learning frame work, to train an ensemble of random forests to predict 3D video saliency. More details regarding *LBVS3D* can be found in [38]. Fig. 3 demonstrates an example of saliency detection from stereoscopic video using 3D VAM of *LBVS3D*.

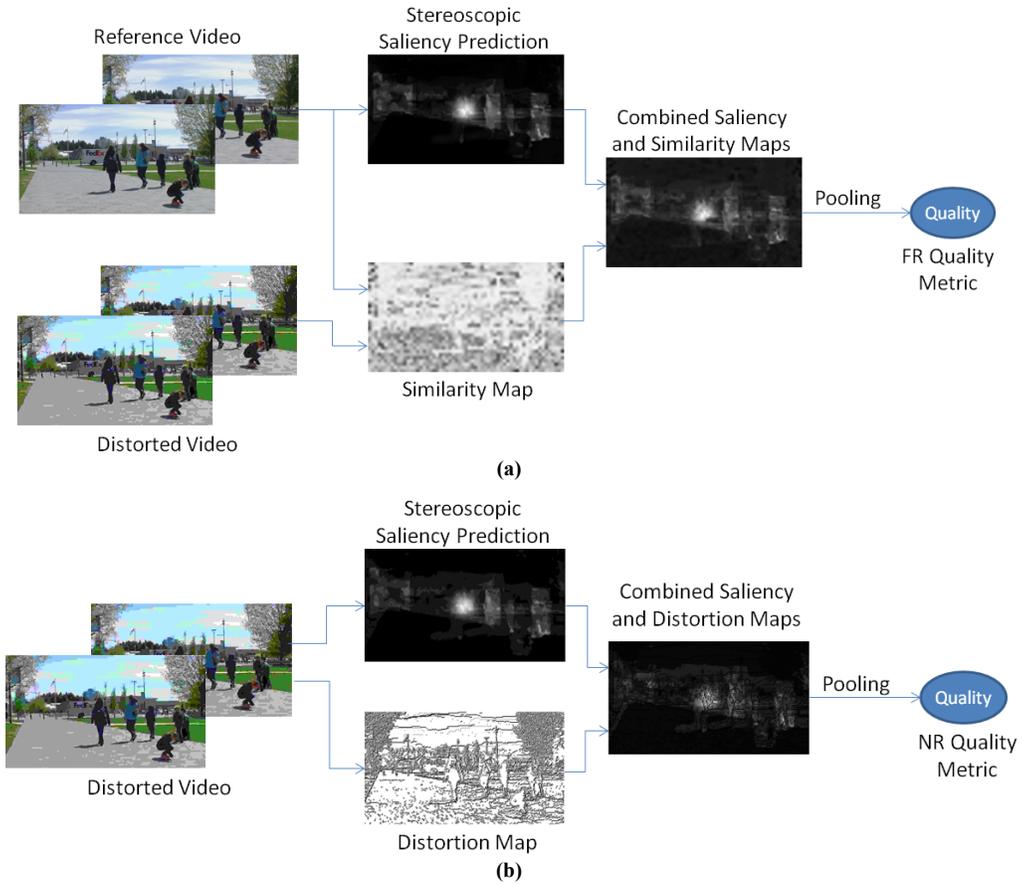

Fig. 1. Saliency inspired quality assessment for stereoscopic video: (a) Full-Reference (FR) and (b) No-Reference (NR) case

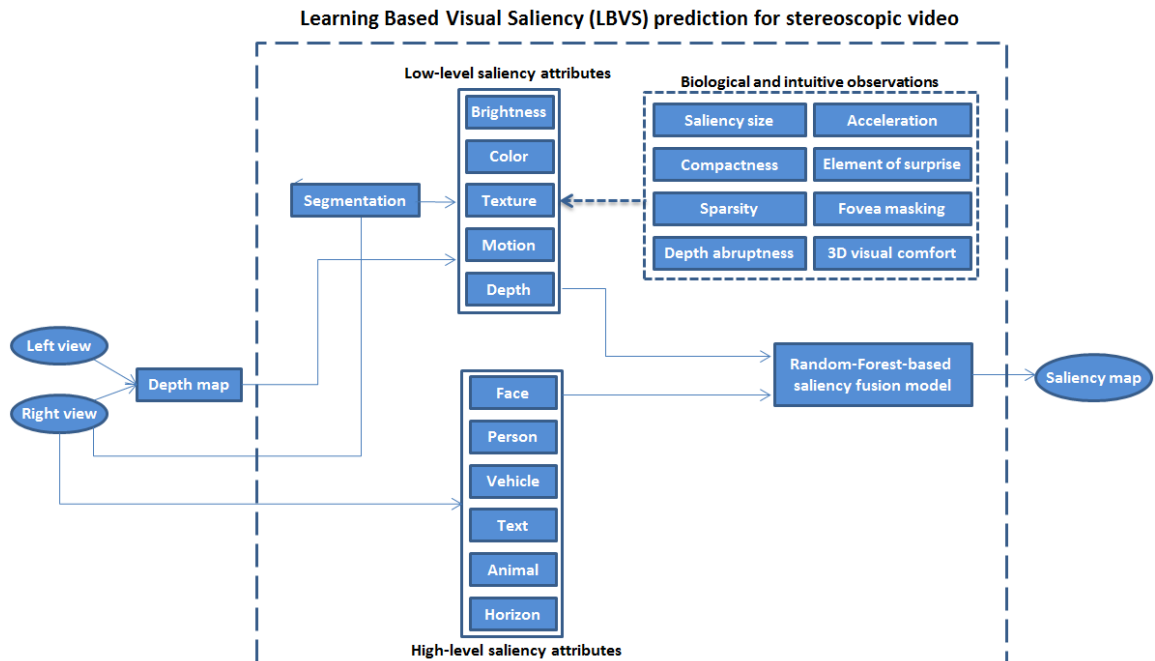

Fig. 2. Block-diagram of *LBVS3D* VAM from [38]

*B. Integration of saliency maps into FR quality metrics*

Visual attention models provide a saliency map for each frame of a video. To be able to use the saliency prediction results in video quality assessment, we only consider those video quality metrics which produce (per frame) a map of similarities, distortions, transform coefficients, errors, or in general the ones for which it is possible to apply the saliency maps as weighting masks. Note that in the case of FR quality evaluation, saliency maps are generated using the reference stereo pair. In this paper, we integrate 3D VAM of LBVS3D into the following FR 3D quality metrics: *Ddl1* (3D quality metric based on Disparity differences) [42], *OQ* (Objective 3D Quality metric for 3D) [43], *CIQ* (Cyclopean view based Image Quality index) [44], *PHVS3D* (Perceptual HVS based 3D metric) [45], *PHSD* (Modified *PHVS3D*) [46], *MJ3D* (3D quality metric proposed by Ming-Jun Chen et. al) [47], *Q_Shao* (3D Quality metric by Shao et. al) [48], *HV3D* (Human Visual system based 3D quality metric for stereo video) [49], and *FLOSIM3D* (Flow based Similarity measure for 3D video) [77]. In addition, we follow what is considered to be a common practice in 3D quality evaluation by using *PSNR* (Peak Signal to Noise Ratio), *SSIM* (Structural SIMilarity) [31], *MS-SSIM* (Multi-Scale *SSIM*) [50], and *VIF* (Visual Information Fidelity) [51] for FR metric integration. In the case of 2D metrics, the overall 3D quality is resulted from averaging the frame qualities in the two views. The rest of this sub-section elaborates on saliency integration for various FR metrics.

*1) PSNR:*

*PSNR* is calculated based on the Mean Square Error (*MSE*) as:

$$PSNR = E_t \left\{ 10\log\left(\frac{255^2}{MSE(t)}\right) \right\} \quad (1)$$

We modify the *MSE* based on saliency maps as follows:

$$MSE_S = E_{x,y}\left\{ |I(x,y) - I'(x,y)|^2 \times S(x,y) \right\} \quad (2)$$

where $x$ and $y$ are pixel coordinates, $t$ denotes the frame number, $I$ and $I'$ are reference and distorted frames, $S$ is the normalized saliency map, and $E_{x,y}$ and $E_t$ denote spatial and temporal mean operators, respectively. Note that *MSE* and *PSNR* are calculated for left and right views separately (using the same saliency map), and the average *PSNR* is considered for each stereo pair.

*2) SSIM [31]:*

Saliency based *SSIM* is computed as:

$$SSIM_S = E_t\left\{ E_{x,y}\left\{ SSIM(I(x,y,t), I'(x,y,t)) \times S(x,y,t) \right\} \right\} \quad (3)$$

where $SSIM(I(x,y,t), I'(x,y,t))$ is the local structural similarity value at pixel location *(x,y)* and time *t*. Similar to *PSNR*, the *SSIM* values are also calculated for each view separately and then averaged for the pair.

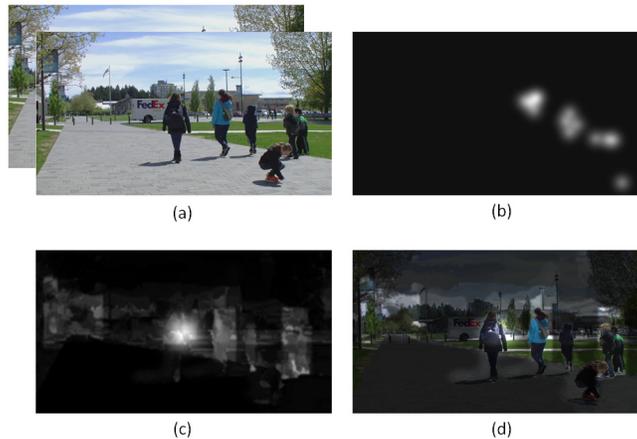

**Fig. 3.** Saliency prediction on stereoscopic video: (a) Original video, (b) eye fixation maps from eye tracking experiments, (c) saliency prediction using LBVS3D method [38], and (d) saliency map super imposed on the original video

*3) MS-SSIM [50]:*

Multi-Scale *SSIM* evaluates structural distortions for a pair of reference-distorted images at a number of scales [50]. In order to apply saliency prediction to *MS-SSIM*, we generate the saliency maps at each scale independently. Then saliency inspired *MS-SSIM* is evaluated as follows:

$$MSSSIM_S = E_t \left\{ E_{x,y} \left\{ l_M(x,y,t) \prod_{m=1}^{M} (c_m(x,y,t) st_m(x,y,t) \times S_m(x,y,t)) \right\} \right\} \quad (4)$$

where $l_m$, $c_m$, and $st_m$ assess luminance, contrast, and structural distortions at scale $m$, $S_m$ is the generated saliency map at scale $m$, and $M$ is the number of decomposition scales. Since *MS-SSIM* is a 2D metric, the saliency based *MS-SSIM* is evaluated for each view separately and the average is calculated as the overall index.

*4) VIF [51]:*

Visual Information Fidelity index evaluates the ratio of visual information between the reference and distorted images [51]. Pixel-wise implementation of *VIF* is defined as follows:

$$VIF_S = E_t \left\{ \frac{\sum_{m=1}^{M} \left\{ \frac{1}{2} \sum_i \sum_k \log\left(1 + \frac{g^2 s_i^2 \lambda_k}{\sigma_v^2 + \sigma_n^2}\right) \times S_m(i,k) \right\}}{\sum_{m=1}^{M} \left\{ \frac{1}{2} \sum_i \sum_k \log\left(1 + \frac{s_i^2 \lambda_k}{\sigma_n^2}\right) \times S_m(i,k) \right\}} \right\} \quad (5)$$

where $M$ is the number of decomposition subbands, $S_m$ is saliency map generated at the size of subband $m$, $i$ and $k$ are spatial subband indices, and $s_i$, $\lambda_k$, $\sigma_n$, and $\sigma_v$ are *VIF* parameters (See [51] for the details).

*5) Ddl1 [42]:*

This metric performs quality assessment on stereoscopic 3D images by weighting the structural similarity values of each view according to their Euclidean disparity differences. Here, we adjust this weighting by incorporating saliency values as follows:

$$Ddl1_{S-left} = E_{x,y} \left\{ SSIM(I_L(x,y,t), I'_L(x,y,t)) \times \left(1 - \frac{\sqrt{D_L(x,y,t)^2 - D'_L(x,y,t)^2}}{255}\right) \times S(x,y,t) \right\}$$

$$Ddl1_{S-right} = E_{x,y} \left\{ SSIM(I_R(x,y,t), I'_R(x,y,t)) \times \left(1 - \frac{\sqrt{D_R(x,y,t)^2 - D'_R(x,y,t)^2}}{255}\right) \times S(x,y,t) \right\} \quad (6)$$

$$Ddl1_S = E_t \left\{ Ddl1_{S-left}(t) + Ddl1_{S-right}(t) \right\}$$

where $D$ and $D'$ are disparity maps, and $L$ and $L'$ denote the left and right view frames, respectively.

*6) OQ [43]:*

You et al. modeled the full-reference stereo quality as the combination of disparity map quality and quality of the views. They used Mean Absolute Differences (*MAD*) to measure the changes in the disparity map and *SSIM* for view quality measurement [43]. In order to modify this metric according to saliency map values, we apply the saliency map as a weighting factor, both in disparity map and view quality evaluations:

$$OQ_S = E_t \left\{ a . IQ_S^d(t) + b DQ_S^e(t) + c . IQ_S^d(t) . DQ_S^d(t) \right\} \quad (7)$$

where

$$IQ_S(t) = SSIM_S(t)$$
$$DQ_S(t) = E_{x,y} \left\{ |D(x,y,t) - D'(x,y,t)| \times S(x,y,t) \right\} \quad (8)$$

and *a, b, c, d*, and *e* are constant coefficients which are derived based on subjective experiments. To conduct a fair comparison, we use the same constant values reported in [43].

*7) CIQ [44]:*

Chen et al. proposed a FR quality assessment framework for stereo images based on generating cyclopean view (fusion of the left and right views in a single image) pictures for reference and distorted views. Then, the overall quality is evaluated as the *SSIM* between the cyclopean pictures from reference and distorted pairs. Saliency inspired *CIQ* is formulated as:

$$CIQ_S = E_t \{ E_{x,y} \{ SSIM(CI(x,y,t), CI'(x,y,t)) \times S(x,y,t) \} \} \quad (9)$$

where *CI* and *CI'* denote the cyclopean view images from the reference and distorted signals.

*8) PHVS3D [45]:*

This metric takes into account the *MSE* of 3D block structures between the reference and distorted stereo pairs [45]. Saliency based *PHVS3D* is defined by:

$$PHVS3D_S = E_t \left\{ 10 \log \left( \frac{255^2}{MSE_{3D-S}(t)} \right) \right\} \quad (10)$$

$$MSE_{3D-S}(t) = \frac{16}{H.W} \sum_{x=1}^{H-3} \sum_{y=1}^{W-3} MSE(A_{xy}(t) - B_{xy}(t)).C_{4\times4}^2.S(x,y,t) \quad (11)$$

where *H* and *W* denote the height and width of the image, $A_{xy}$ and $B_{xy}$ are *3D-DCT* coefficients for the reference and distorted views, and *C* is a Contrast Sensitivity Function (*CSF*) mask [45].

*9) PHSD [46]:*

This metric is an improved version of *PHVS3D* [45] (mentioned above) which considers the *MSE* between the depth maps in conjunction with the *MSE* of block structures [46]. Our modification of *PHSD* is in two levels of *MSE*, in both block structures and depth maps, and is formulated by:

$$PHSD_S(t) = E_t \left\{ 10 \log \left( \frac{255^2}{(1-\varepsilon).MSE_{i-S} + \varepsilon.MSE_{d-S}} \right) \right\} \quad (12)$$

where

$$MSE_{d-S} = E_t \{ E_{x,y} \{ |D(x,y,t) - D'(x,y,t)|^2 \times S(x,y,t) \} \}$$
$$MSE_{i-S} = E_t \{ E_{x,y} \{ MSE_{bs}(x,y) . \frac{MSE_{bs}(x,y)}{MSE_{bs}(x,y) + \alpha \sigma_d^2(x,y)} \times S(x,y,t) \} \} \quad (13)$$

and $\sigma_d^2$ is the variance of depth map at spatial location *(x,y)* and $MSE_{bs}$ is the error value calculated for 3D block structures [46].

*10) MJ3D [47]:*

In this approach, Multi Scale *SSIM* is utilized for quality assessment of cyclopean view pictures constructed from reference and distorted stereo pairs [47]. We modify *MJ3D* based on saliency maps as follows:

$$MJ3D_S = E_t \left\{ E_{x,y} \left\{ l_{CI-M}(x,y,t) . \prod_{m=1}^{M} (c_{CI-m}(x,y,t) st_{CI-m}(x,y,t) \times S_m(x,y,t)) \right\} \right\} \quad (14)$$

where $l_{CI}$, $c_{CI}$, and $st_{CI}$ are the luminance, contrast, and structural components of *MS-SSIM* in each scale, generated from the cyclopean view images.

*11) Q_Shao [48]:*

Shao and colleagues proposed a quality assessment method for stereo images which is based on image regions classification. In this method, each view (in both reference and distorted stereo pairs) is partitioned to three possible categories: non-corresponding (*nc*), binocular fusion (*bf*), and binocular suppression (*bs*) regions. Three quality components are calculated based on the three regions and combined into an overall index as follows [48]:

$$Q = E_t \{\omega_{nc} Q_{nc} + \omega_{bf} Q_{bf} + \omega_{bs} Q_{bs}\} \qquad (15)$$

where $w_{nc}$, $w_{bf}$, and $w_{bs}$ are weighting coefficients for each quality component. Each component is computed as the average of per-pixel values over the corresponding region. Saliency information is therefore incorporated in each of the components, as a weighting factor to emphasize the visually important image regions.

*12) HV3D [49]:*

Human Visual system based 3D video quality metric (*HV3D*) evaluates the perceived 3D quality of stereoscopic videos as a combination of depth map quality and quality of the views. *HV3D* is formulated as a combination of three terms [49]:

$$HV3D = E_t \left\{ \left( \sum_{i=1}^{N} \frac{SSIM(IDCT(XC_i), IDCT(XC_i'))}{N} \right)^{\beta_1} \cdot (VIF(D, D'))^{\beta_2} \cdot \left( \sum_{i=1}^{N} \frac{\sigma_{d_i}^2}{N \cdot \max(\sigma_{d_j}^2 \mid j=1,2,...,N)} \right)^{\beta_3} \right\} \qquad (16)$$

where $XC_i$ is the cyclopean-view model for the $i^{th}$ matching block pair in the reference 3D view, $XC_i'$ is the cyclopean-view model for the $i^{th}$ matching block pair in the distorted 3D view, *IDCT* stands for inverse 2D discrete cosine transform, *N* is the total number of blocks in each view, $\beta_1$, $\beta_2$ and $\beta_3$ are constant exponents, and $\sigma_{di}^2$ is the local variance of block *i* in the disparity map of the 3D reference view. In order to incorporate the saliency information in each of the three quality components of *HV3D*, we use the saliency based $SSIM_S$ index (see (3)) for the first component and the $VIF_S$ index (see (5)) for the second component. For the third component (variances of the blocks), a single average saliency value (average of saliency map values) for each block is used as a weight for the variance term. The same constant parameter values are used as the ones reported in [49].

*13) FLOSIM3D [77]:*

*FLOSIM3D* exploits the 3D video quality by combining temporal, spatial, and depth quality attributes followed by a pooling strategy. The temporal component (noted by $Q^i_{FL\text{-}l}$ and $Q^i_{FL\text{-}r}$ for *i-th* left & right view frames) is computed by measuring dispersion of mean, variance, and minimum eigen value of patches between a reference frame and corresponding distorted frame. Spatial component (noted by $Q^i_{sl}$ and $Q^i_{sr}$ for left & right views) is a dissimilarity measure of (*1-MS-SSIM* [50]) over corresponding views and frames. And depth component (noted by $Q^i_{dl}$ and $Q^i_{dr}$ for left and right view depth maps) is a dissimilarity measure of (*1-MS-SSIM* [50]) over corresponding depth maps. *FLOSIM3D* is formulated as follows:

$$FLOSIM = \frac{1}{2T} \sum_{i=1}^{T} (Q^i_{sl} Q^i_{FL-l} + Q^i_{sr} Q^i_{FL-r})$$

$$Q_d = \frac{1}{2T} \sum_{i=1}^{T} (Q^i_{dl} + Q^i_{dr}) \qquad (17)$$

$$FLOSIM_{3D} = FLOSIM \times Q_d$$

where *T* is the total number of frames. In order to integrate 3D saliency maps to the formulation of *FLOSIM3D*, we apply the masks to the spatial and depth components of this metric. As both of these components utilize *MS-SSIM* [50], we can leverage the already integrated *MS-SSIM* of section *II.B.3* (see above) and follow the same methodology here. Once saliency maps are applied to *MS-SSIM* of depth and spatial components, the new *FLOSIM3D$_S$* metric will consider saliency weights for quality measurements.

*C. Integration of saliency maps into NR quality metrics*

No reference quality assessment is generally a much more difficult task than the full reference quality assessment as no information is available about the reference data. As a consequence, NR quality metrics usually aim at evaluating the quality when only a specific type of distortion is present. Due to widespread application of image and video compression, many NR quality metrics have been proposed so far that assess the sharpness, blurriness, or blockiness of images or videos. Compared to FR 3D video quality metrics, there is less number of NR 3D metrics proposed in the literature. Here, we use *QA3D* (Quality Assessment metric for 3D) [52], *NOSPDM* (NO reference Stereoscopic Parallax based Distortion Metric) [29], *Q_Ryu* (3D quality metric by Ryu et. al) [30], *and APT* (Auto-regressive Plus Threshold) [78] for saliency integration, as these are NR 3D metrics that can be modified according to the available saliency maps. In addition to NR 3D metrics, we also apply our saliency maps to several other 2D metrics, which include: *IQVG* (Image Quality index based on Visual saliency guided sampling and Gabor filtering) [25], *GBIM* (Generalized Block-Edge Impairment Metric) [53], *NRPBM* (No Reference Perceptual Blur Metric) [54], *Q_blur_Farias* (blur quality metric by Farias et. al) [26], *Q_block_Farias* (blockiness quality metric by Farias et. al) [26], *Q_Sadaka* (Quality metric by Sadaka et. al) [27], *VQSM* (Visual Quality Saliency based Metric) [28], and *AQI* (Anisotropy Quality Index) [55]. In the case of 2D metrics, the overall quality is measured as the average quality of the frames for the two views. Note that in the case of NR quality assessment, saliency maps are generated using the distorted stereo pair as no reference is available. This requires accurate saliency prediction from distorted videos. *LBVS3D* is capable of efficiently detect the salient regions in a video, even in the presence of distortions. Fig. 4 demonstrates examples of distorted video frames and how saliency maps are extracted using the *LBVS3D*. The rest of this sub-section elaborates on saliency integration for each metric.

*1) IQVG [25]:*

Image Quality index based on Visual saliency guided sampling and Gabor filtering, *IQVG*, performs blind quality assessment of 2D images by applying Support Vector Regression (SVR) on features extracted from sampled image patches [25]. These patches are selected based on 2D saliency information. Here, we swap the 2D saliency maps used in *IQVG* with the stereo saliency maps of *LBVS3D*. The rest of the process remains unchanged.

*2) GBIM [53]:*

Generalized Block-Edge Impairment Metric, *GBIM*, measures blockiness artifacts present in digital video and coded images [53]. Blockiness across horizontal and vertical edges are averaged to formulate the *GBIM* as:

$$GBIM = E_t \left\{ \frac{M_h + M_v}{2 \times E} \right\} \quad (18)$$

where $E$ is the average inter-pixel difference, and $M_h$ and $M_v$ measure the horizontal and vertical edge blockiness. 3D saliency based $M_h$ is defined by:

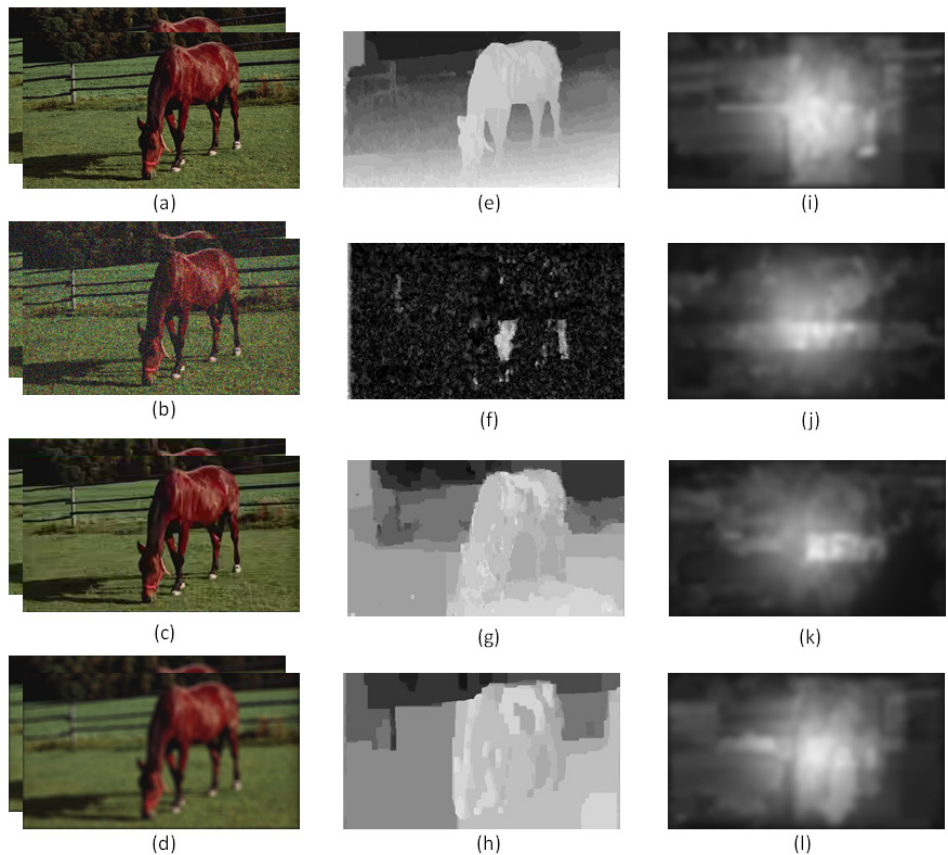

**Fig. 4. Effect of distortion on depth map generation and saliency prediction; video frames: (a) reference video, (b) Additive White Gaussian Noise (AWGN), (c) 3D video compression, and (d) Gaussian blur, generated depth maps: (e), (f), (g), and (h), and predicted saliency maps using LBVS-3D method [38]: (i), (j), (k), and (l).**

$$M_{h-S} = E_{x,y}\{W(x,y).D_cF(x,y).S(x,y)\} \quad (19)$$

where $W$ is a diagonal weighting matrix and $D_cF$ is the inter-pixel difference between each of the horizontal block boundaries [53]. $M_{v-S}$ is defined similarly.

*3) NRPBM [54]:*

Roffet et al. proposed a blurriness metric for 2D images based on the average horizontal and vertical block edge differences in the reference and distorted pictures [54]. We use 3D saliency map values as weights to these differences towards the overall index as follows:

$$NRPBM_S = E_t\left\{1 - Max\left(\frac{\sum_{x,y} DV_{ver}(x,y,t) \times S(x,y,t)}{\sum_{x,y} DF_{ver}(x,y,t) \times S(x,y,t)}, \frac{\sum_{x,y} DV_{hor}(x,y,t) \times S(x,y,t)}{\sum_{x,y} DV_{ver}(x,y,t) \times S(x,y,t)}\right)\right\} \quad (20)$$

where $DV$ and $DF$ are the differences between the original and blurred image, in vertical and horizontal directions [54].

*4) Q_Farias [26]:*

Farias and Akamine proposed two NR metrics for quality assessment of 2D images. These metrics measure the blockiness and blurriness of an image [26]. In this approach, the blurriness metric is defined as the average of edge width. And the blockiness metric is defined by the average horizontal and vertical differences. Saliency based versions of the mentioned metrics are formulated as follows:

$$Blur_S = E_t\{E_{x,y}(width(x,y,t) \times S(x,y,t))\} \quad (21)$$

$$Block_S = E_t\left\{\frac{1}{H.W}\left(\frac{\sum_{i=1}^{H/8}\sum_{j=1}^{W} D_v^8(i,j,t) \times S(x,y,t)}{\sum_{i=1}^{H}\sum_{j=1}^{W} D_v(i,j,t) \times S(x,y,t)} + \frac{\sum_{i=1}^{H}\sum_{j=1}^{W/8} D_h^8(i,j,t) \times S(x,y,t)}{\sum_{i=1}^{H}\sum_{j=1}^{W} D_h(i,j,t) \times S(x,y,t)}\right)\right\} \quad (22)$$

where $H$ and $W$ are height and width of the image, $D_h$ and $D_v$ are the horizontal and vertical brightness differences, and $D_h^8$ and $D_v^8$ are brightness differences at the borders of 8×8 block partitions.

*5) Q_Sadaka [27]:*

Sadaka et al. designed an image sharpness metric based on the Just Noticeable Blur (JNB) and 2D image saliency maps [27]. In this method, the sharpness metric is defined as:

$$M = E_t\left\{\left(\sum_{R \in I}\left|D_R \cdot \frac{\sum_{i \in R} S(x,y,t)}{\sum_{i \in I} S(x,y,t)}\right|^{\beta}\right)^{\frac{-1}{\beta}}\right\} \quad (23)$$

where $D_R$ indicates the amount of perceived blur in the region $R$ [27]. Here, we substitute the 2D image saliency maps used in (23) with our 3D video saliency maps, which are proven to provide superior stereo saliency detection [38].

*6) VQSM [28]:*

Visual Quality Saliency based Metric (*VQSM*) is a 2D NR image quality metric which measures the sharpness and smoothness of an image [28]. We modify the sharpness and smoothness components using the available 3D saliency information as follows:

$$Q_{sh} = E_{x,y}\left\{\sqrt{g_x^2(x,y,t) + g_y^2(x,y,t)} \times S(x,y,t)\right\} \quad (24)$$

$$Q_{sm} = E_{x,y}\{\sigma_I^{5\times 5} \times \overline{S}^{5\times 5}\} \quad (25)$$

where $g_x$ and $g_y$ are the gradient values of image brightness component in horizontal and vertical directions, $\sigma_I^{5\times 5}$ is standard deviation over a *5×5* window centered at *(x,y)*, and $\overline{S}$ is the average saliency values over the same window. The overall quality

index is calculated using the same method as *VQSM*, and averaged over the frames:

$$VQSM_S = E_t\{\alpha_1 Q_{sh}^2 + \alpha_2 Q_{sh} + \alpha_3 Q_{sm}^2 + \alpha_4 Q_{sm} + \alpha_5\} \quad (26)$$

*7) AQI [55]:*

Anisotropy Quality Index (*AQI*) is a 2D blind image quality metric based on measuring the variance of the expected entropy of an image upon a set of predefined directions [55]. We incorporate saliency probabilities of pixels into the entropy used by *AQI* index as weighting coefficients.

*8) QA3D [52]:*

*QA3D* is a NR video quality metric for stereoscopic images which is designed to assess the transmission artifacts (blockiness, sharpness, and edginess) [52]. In this method, first a hard threshold is applied to the disparity maps of the reference and distorted pair to set the disparity values smaller than the threshold to zero. Then, a disparity index for frame *n* is defined by:

$$D_n = E_{x,y}\{D(x,y)\} \quad (27)$$

The disparity index is used to define a dissimilarity index between the views:

$$S_m = \frac{1}{10}\left(\sum_{i=n-p}^{n-1} D_i - D_n P\right) \times D_n \quad (28)$$

where *p* is the number of previous frames used. This index along with an edge based difference measure ($D_E$) form the overall *QA3D*:

$$QA3D = E_t\left\{1 - \frac{S_m + D_E}{2}\right\} \quad (29)$$

Saliency information is incorporated in this metric both in the dissimilarity index $S_m$ and difference index $D_E$ as spatial weighting coefficients.

*9) NOSPDM [29]:*

Gu et al. proposed a parallax compensation based distortion metric (*NOSPDM*) for JPEG compressed stereoscopic images [29] defined as (for each stereo pair):

$$NOSPDM = (2-\mu_R)QJPEG_L + \mu_R QJPEG_R - \lambda \max\{QJPEG_L, QJPEG_R\} + \cos^{-1}\left(\frac{L.R}{\|L\|_2\|R\|_2}\right) + \omega_s.\cos^{-1}\left(\frac{S_L.S_R}{\|S_L\|_2\|S_R\|_2}\right) \quad (30)$$

where *L* and *R* are the two view images, $S_L$ and $S_R$ are 2D saliency maps, $\mu_R$, $\omega_S$, and $\lambda$ are constant parameters, and *QJPEG* measures the sharpness of each view as follows:

$$QJPEG = \alpha + \beta\left(\frac{B_h + B_v}{2}\right)^{\gamma_1}\left(\frac{A_h + A_v}{2}\right)^{\gamma_2}\left(\frac{Z_h + Z_v}{2}\right)^{\gamma_3} \quad (31)$$

where *α, β, γ₁, γ₂, γ₃* are constant parameters, $B_h$ and $B_v$ are blockiness across horizontal and vertical edges, $A_h$ and $A_v$ are the average absolute difference between in-block image samples in horizontal and vertical directions, and $Z_h$ and $Z_v$ are the horizontal and vertical zero crossing rates [29]. We use the 3D saliency information in the *QJPEG* terms for the two views in each of the three components as weights to the pixel values and zero crossings.

*10) Q_Ryu [30]:*

Ryu and Sohn proposed a NR quality metric for stereoscopic images which takes into account blurriness and blockiness of an image pair [30]. In this approach, a pair of blurriness and blockiness maps are generated for each view, and combined with 2D saliency maps generated from the views. We substitute the 2D saliency maps used in this approach with our 3D saliency maps.

*11) APT [78]*

AR-plus thresholding (*APT*) is a no-reference blind quality metric designed for quality assessment of 3D synthesized DIBR (depth image-based rendering) images. *APT* computes two binary maps for measuring geometrical ($M_d$) and non-geometrical natural ($M_r$) image distortions. To generate the geometric binary distortion map, *APT* already incorporates saliency maps from FES VAM [79] in a thresholding operation to extract γ % of most salient regions. The overall metric is formulated as follows (for a stereo-pair):

$$Q = \frac{1}{L}\sum_{l=1}^{L}\left(\frac{2M_d(l).M_r(l)+\varepsilon}{M_d(l)^2 + M_r(l)^2 + \varepsilon}\right)^{\alpha} \quad (32)$$

where $l$ is the pixel index, $L$ is the number of pixels in the image, $\varepsilon$ is a constant used for stability, and $\alpha$ is the Minkowski exponent. We integrate 3D saliency in *APT* by substituting 3D saliency maps with the FES maps already used. All implementation parameters are selected based on the information provided in [78] paper.

## III. EXPERIMENTS

We modify the FR and NR quality metrics described in Section II using stereo saliency information and evaluate the performance of the modified metrics in comparison to the original metrics. This section reviews the incorporated video database in the experiments, as well as the subjective tests procedure.

*A. Stereoscopic video database*

Sixteen stereo videos were chosen from three different datasets for our subjective experiments. These videos were taken from: 1) sequences provided by MPEG for video compression standardization activities. These sequences were originally captured in multi views. We select only two views, according to MPEG recommendations for subjective tests for video compression studies [56]. 2) Digital Multimedia Lab (DML) stereo video sequences at the University of British Columbia (available online [57]), and 3) test videos used in [46]. Table I contains the description of the stereo video database used in our experiments.

When introducing video databases, it is a common practice to measure the spatial and temporal complexity of the videos in the database to ensure the videos are from dynamic scenes with a wide range of spatial and temporal complexity [58]. Fig. 5 shows the distribution of the video database used in our experiments. It is observed from Fig. 5 that videos cover a wide range of temporal and spatial complexities. In addition to spatial and temporal video complexities, depth bracket of the scenes is measured and reported in Table I. The depth bracket (or range) is a rough estimate of the distance between the closest and the farthest visually important objects in each scene [59]. Since the camera information (coordinates, focal length, and fundamental matrix) are not available for all sequences, disparity maps are first converted to depth maps using the method reported in [38] and then depth differences are measured. Visually important objects are selected based on the available 3D saliency maps.

In order to evaluate the performance of various quality metrics, several different types of distortions and artifacts are simulated over our video database. Table II contains the details regarding the distortions. More details can be found in [49]. Note that after applying the distortions, a new disparity map is generated for each stereo video using the MPEG Depth Estimation Reference Software (DERS) [60]. Also, capturing artifacts such as window violation, vertical parallax, depth plane curvature, keystone distortion, or shear distortion are not considered. Using 7 different distortions (at 13 levels) for 16 reference videos resulted in 208 stereo videos in our database.

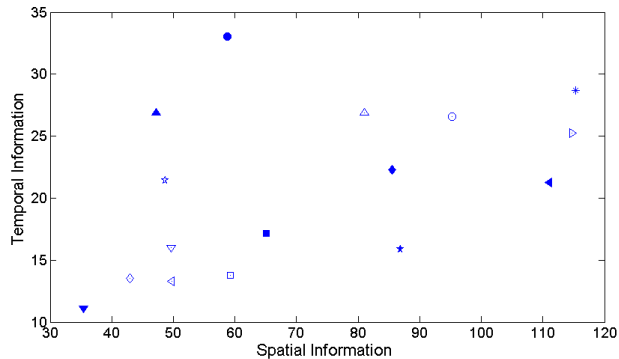

Fig. 5. Spatial and temporal information distribution for the stereo video database used in our experiments

*B. Subjective tests*

Subjective experiments were conducted using our stereo video database with participation of 88 subjects. Test material was presented to the viewers using a HD 3D TV (Hyundai S465D) with passive glasses. Video presentation was performed using the Single Stimulus (SS) method in the accordance to ITU BT. 500-13 [61]. Details regarding the subjective tests can be found in [49].

## IV. RESULTS AND DISCUSSIONS

In this study, saliency maps from *LBVS3D* (and 4 other different 3D VAMs) are incorporated to the formulation of 13 full-reference and 12 no-reference objective quality metrics. In order to understand the impact of saliency integration in the performance of these metrics, we need to evaluate their performance with and without the saliency integration. To this end, subjective evaluations are performed to measure the FR and NR performance of objective metrics, with and without saliency information. Following the common practice in the quality assessment literature, we measure the metrics performances using *PCC*, *SCC*, *RMSE*, and *OR* objective measures. This section elaborates on the specifics of this process.

After collecting the subjective quality scores, an overall Mean Opinion Score (*MOS*) is calculated by averaging the individual subjective scores and removing the outliers [49]. An objective metric value is also computed using each of the FR and NR metrics to compare with the subjective results. Four different performance metrics are used in our analyses: 1,2) Pearson Linear Correlation Coefficient (*PLCC* or *PCC*) and Root Mean Square Error (*RMSE*) which measure the accuracy of each objective metric in predicting MOS values. 3) Spearman Correlation Coefficient (*SCC*) which evaluates the monotonicity of a mapping between an objective metric and *MOS* results, 4) Outlier Ratio (*OR*), to measure the consistency of objective metrics in prediction of *MOS* values.

Performance of objective FR and NR metrics are calculated for our stereo video database using the mentioned performance metrics and compared with that of the same metrics when stereo saliency maps are taken into account. Table III and Table IV show the performance of FR and NR metrics for the original metrics and when the stereo saliency information is incorporated. It is observed from these tables that saliency prediction in general improves the performance of both objective FR and NR metrics. In the case of FR metrics, the improvements are on average less than the NR case. This is due to the fact that information from the reference video is available for FR quality assessment and thus more accurate assessment is possible. It is worth noting that some of the NR metrics evaluated in our study incorporate 2D saliency maps in their original design. These metrics (*IQVG* [25], *Q_Farias* [26], *Q_Sadaka* [27], *VQSM* [28], *NOSPDM* [29], *Q_Ryu* [30], *APT* [78]) receive less improvement after being modified by stereo saliency information.

**Table 1** Stereoscopic video database

| Sequence | Resolution | Frame Rate (fps) | Number of Frames | Spatial Complexity (Spatial Information) | Temporal Complexity (Temporal Information) | Depth Range (cm) |
|---|---|---|---|---|---|---|
| **Poznan_Hall2** | 1920×1088 | 25 | 200 | Low (35.4658) | Low (11.1460) | High (28.93) |
| **Undo_Dancer** | 1920×1088 | 25 | 250 | High (81.0423) | High (26.9021) | High (30.69) |
| **Poznan_Street** | 1920×1088 | 25 | 250 | High (95.3103) | High (26.5562) | High (34.01) |
| **GT_Fly** | 1920×1088 | 25 | 250 | Medium (58.8022) | High (33.0102) | High (31.02) |
| **Cokeground** | 1000×540 | 30 | 210 | High (86.9096) | Medium (15.9128) | Low (4.99) |
| **Ball** | 1000×540 | 30 | 150 | Medium (49.7701) | Low (13.3074) | Medium (15.53) |
| **Kendo** | 1024×768 | 30 | 300 | Medium (47.2172) | High (26.8791) | High (21.39) |
| **Balloons** | 1024×768 | 30 | 500 | Medium (48.6726) | High (21.4660) | Low (5.84) |
| **Lovebird1** | 1024×768 | 30 | 300 | Medium (59.2345) | Low (13.8018) | Medium (15.01) |
| **Newspaper** | 1024×768 | 30 | 300 | High (65.1173) | Medium (17.1297) | Low (5.09) |
| **Soccer2** | 720×480 | 30 | 450 | High (115.2781) | High (28.6643) | Medium (16.99) |
| **Alt-Moabit** | 512×384 | 30 | 100 | High (111.0437) | High (21.2721) | Medium (13.36) |
| **Hands** | 480×270 | 30 | 251 | High (114.6755) | High (25.2551) | Medium (15.86) |
| **Flower** | 480×270 | 30 | 112 | Medium (43.0002) | Low (13.5305) | Low (5.86) |
| **Horse** | 480×270 | 30 | 140 | High (85.4988) | High (22.3184) | Medium (13.56) |
| **Car** | 480×270 | 30 | 235 | Medium (49.6162) | Medium (16.0197) | High (24.21) |

**Table II** Different types of distortions

| Artifact / Distortion | Description | Parameters | Affects views separately | Affects both views simultaneously |
|---|---|---|---|---|
| **AWGN** | Additive White Gaussian Noise | zero mean and variance value 0.01 | X | |
| **Blur** | GLPF: Gaussian Low Pass Filter | size 4 and the standard deviation of 4 | X | |
| **Intensity Shift** | Increased brightness values | Increment by 20 (out of 255) | X | |
| **Simulcast Coding** | Simulcast compression of the views | HEVC HM 16.7 [65], GOP 4, QP 35, 40, Low Delay configuration profile | X | |
| **Disparity Map Compression** | Synthesizing views using a highly compressed disparity map | HEVC HM 16.7 [65], GOP 4, QP 25, 45, Low Delay configuration profile | | X |
| **3D Video Compression** | 3D video compression | HEVC based 3D HTM 9 [65], GOP 8, QP 25, 30, 35, 40, Random Access High Efficiency profile | | X |
| **View Synthesis** | Synthesizing one view | Using VSRS 3.5 [66] for synthesizing one view based on disparity map and the other view | | X |

Other observations from Table III and IV:
- *PCC*, *SCC*, and *RMSE* consistently show slight improvements when incorporating saliency. However, Outlier Ratio results are not as conclusive.
- *PSNR* receives 3.5% *PCC* improvement when integrated with 3D saliency. This is the highest amount of improvement compared to the other FR metrics. Note that it was expected that *PSRN* receives a big improvement as it does not perform any smart comparisons but to subtract the pixel values.
- Within NR metrics, *NRPBM* [54] receives the highest *PCC* improvement of 6.5%, which is considered as a quite significant improvement.

In addition to incorporating stereo saliency information generated using the *LVBS3D* VAM in the FR and NR quality assessment tasks, we examine the added value of saliency information resulted from several other state-of-the-art 3D VAMs namely 3D VAMs of Fang et al. [37], Coria et al. [62], and Park et al. [63]. Following what is considered to be a common practice in saliency prediction studies, we also add to the evaluations the results from the 2D VAM of Itti et al. [64]. Fig. 6 shows the improvements in quality assessment achieved by using various VAMs. It is observed from Fig. 6 that 3D VAM of *LBVS3D* has resulted in highest improvements in quality assessment. Moreover, NR metrics that already use saliency information receive less improvement compared to other NR metrics.

**Table III.** Statistical performance of different FR quality metrics with and without integration of the saliency maps

| Quality Metric | Performance Metric | PCC | | SCC | | RMSE | | Outlier Ratio | |
|---|---|---|---|---|---|---|---|---|---|
| | | Original | Saliency Inspired | Original | Saliency Inspired | Original | Saliency Inspired | Original | Saliency Inspired |
| PSNR | | 0.6454 | 0.6800 | 0.6350 | 0.6646 | 10.388 | 9.671 | 0.0167 | 0.0083 |
| SSIM [31] | | 0.6844 | 0.7113 | 0.6213 | 0.6946 | 9.852 | 9.057 | 0.0083 | 0.0083 |
| MS-SSIM [50] | | 0.7071 | 0.7219 | 0.7180 | 0.7481 | 9.999 | 9.009 | 0.0083 | 0.0083 |
| VIF [51] | | 0.7257 | 0.7380 | 0.7204 | 0.7349 | 9.166 | 8.947 | 0 | 0 |
| Ddl1 [42] | | 0.7370 | 0.7638 | 0.7321 | 0.7557 | 8.732 | 8.556 | 0 | 0 |
| OQ [43] | | 0.7580 | 0.7709 | 0.7900 | 0.7993 | 8.610 | 8.500 | 0 | 0 |
| CIQ [44] | | 0.7200 | 0.7451 | 0.7080 | 0.7346 | 9.446 | 8.884 | 0.0083 | 0 |
| PHVS3D [45] | | 0.7837 | 0.8022 | 0.8233 | 0.8238 | 8.420 | 8.300 | 0 | 0 |
| PHSD [46] | | 0.7911 | 0.8234 | 0.7841 | 0.8010 | 8.321 | 8.067 | 0 | 0 |
| MJ3D [47] | | 0.8640 | 0.8698 | 0.8947 | 0.9033 | 7.229 | 7.178 | 0 | 0 |
| Q_Shao [48] | | 0.8348 | 0.8524 | 0.7988 | 0.8349 | 7.902 | 7.436 | 0 | 0 |
| HV3D [49] | | 0.9082 | 0.9231 | 0.9130 | 0.9343 | 6.433 | 6.267 | 0 | 0 |
| FLOSIM3D [77] | | 0.7639 | 0.7712 | 0.7511 | 0.7596 | 8.527 | 8.467 | 0 | 0 |

**Table IV.** Statistical performance of different NR quality metrics for each specific type of distortion

| Quality Metric | Performance Metric | PCC | | SCC | | RMSE | | Outlier Ratio | |
|---|---|---|---|---|---|---|---|---|---|
| | | Original | Saliency Inspired | Original | Saliency Inspired | Original | Saliency Inspired | Original | Saliency Inspired |
| IQVG [25] | | 0.6713 | 0.6805 | 0.6892 | 0.6956 | 9.923 | 9.901 | 0.0083 | 0.0083 |
| GBIM [53] | | 0.6065 | 0.6538 | 0.5897 | 0.6251 | 10.849 | 10.102 | 0.0167 | 0.0083 |
| NRPBM [54] | | 0.5980 | 0.6634 | 0.6001 | 0.6678 | 10.963 | 10.038 | 0.0167 | 0.0083 |
| Q_blur_Farias [26] | | 0.6312 | 0.6449 | 0.6229 | 0.6437 | 10.430 | 10.321 | 0.0083 | 0.0083 |
| Q_block_Farias [26] | | 0.6494 | 0.6523 | 0.6550 | 0.6591 | 10.432 | 10.277 | 0.0083 | 0.0083 |
| Q_Sadaka [27] | | 0.6668 | 0.6878 | 0.6790 | 0.6889 | 9.993 | 9.911 | 0.0083 | 0.0083 |
| VQSM [28] | | 0.6903 | 0.7009 | 0.6945 | 0.7178 | 8.987 | 8.690 | 0.0083 | 0 |
| AQI [55] | | 0.6882 | 0.7426 | 0.6721 | 0.7448 | 8.995 | 8.766 | 0.0083 | 0 |
| QA3D [52] | | 0.7127 | 0.7633 | 0.7089 | 0.7467 | 8.680 | 8.012 | 0 | 0 |
| NOSPDM [29] | | 0.7843 | 0.7919 | 0.7911 | 0.7999 | 7.943 | 7.903 | 0 | 0 |
| Q_Ryu [30] | | 0.8475 | 0.8533 | 0.8410 | 0.8557 | 7.687 | 7.559 | 0 | 0 |
| APT [78] | | 0.6747 | 0.6804 | 0.6530 | 0.6599 | 9.346 | 9.141 | 0.0083 | 0.0083 |

**Table V.** PCC values for different FR quality metrics and for each specific type of distortion

| Quality Metric | Additive White Gaussian Noise | | Simulcast Compression | | Blurring | | Brightness Shift | | 3D Video Compression | | View Synthesis | | Depth Map Compression | |
|---|---|---|---|---|---|---|---|---|---|---|---|---|---|---|
| Distortion | Original | Saliency Inspired | Original | Saliency Inspired | Original | Saliency Inspired | Original | Saliency Inspired | Original | Saliency Inspired | Original | Saliency Inspired | Original | Saliency Inspired |
| PSNR | 0.6832 | 0.6894 | 0.7098 | 0.7358 | 0.5919 | 0.6089 | 0.4931 | 0.5101 | 0.7400 | 0.7492 | 0.6393 | 0.6399 | 0.6503 | 0.6592 |
| SSIM [31] | 0.7716 | 0.7809 | 0.7530 | 0.7734 | 0.7159 | 0.7335 | 0.7653 | 0.7867 | 0.6998 | 0.7212 | 0.6539 | 0.6678 | 0.6700 | 0.6832 |
| MS-SSIM [50] | 0.7888 | 0.7892 | 0.7768 | 0.7890 | 0.8209 | 0.8345 | 0.8367 | 0.8445 | 0.7199 | 0.7268 | 0.6234 | 0.6411 | 0.7488 | 0.7637 |
| VIF [51] | 0.7967 | 0.8065 | 0.7694 | 0.7757 | 0.8232 | 0.8403 | 0.8523 | 0.8722 | 0.7307 | 0.7419 | 0.6059 | 0.6099 | 0.7535 | 0.7608 |
| Ddl1 [42] | 0.6228 | 0.6317 | 0.7620 | 0.7821 | 0.7433 | 0.7543 | 0.7761 | 0.7807 | 0.7200 | 0.7276 | 0.7255 | 0.7372 | 0.8132 | 0.8255 |
| OQ [43] | 0.7133 | 0.7178 | 0.6859 | 0.6932 | 0.7042 | 0.7183 | 0.6519 | 0.6605 | 0.7411 | 0.7497 | 0.6816 | 0.6933 | 0.7071 | 0.7111 |
| CIQ [44] | 0.7769 | 0.7834 | 0.7741 | 0.7956 | 0.8325 | 0.8378 | 0.7240 | 0.7289 | 0.7556 | 0.7711 | 0.7123 | 0.7241 | 0.7701 | 0.7922 |
| PHVS3D [45] | 0.6918 | 0.6990 | 0.7960 | 0.8178 | 0.7244 | 0.7307 | 0.6079 | 0.6180 | 0.8194 | 0.8289 | 0.7366 | 0.7460 | 0.7840 | 0.8000 |
| PHSD [46] | 0.6484 | 0.6545 | 0.8522 | 0.8761 | 0.7523 | 0.7644 | 0.6249 | 0.6352 | 0.8330 | 0.8566 | 0.7532 | 0.7677 | 0.8285 | 0.8339 |
| MJ3D [47] | 0.8277 | 0.8333 | 0.8452 | 0.8604 | 0.8123 | 0.8189 | 0.8601 | 0.8678 | 0.8009 | 0.8193 | 0.7219 | 0.7389 | 0.7018 | 0.7149 |
| Q_Shao [48] | 0.7988 | 0.8056 | 0.8233 | 0.8483 | 0.8278 | 0.8360 | 0.7812 | 0.7959 | 0.7923 | 0.8101 | 0.7088 | 0.7253 | 0.7245 | 0.7446 |
| HV3D [49] | 0.7994 | 0.8075 | 0.8312 | 0.8578 | 0.8108 | 0.8178 | 0.8412 | 0.8500 | 0.8965 | 0.9065 | 0.8881 | 0.9005 | 0.8603 | 0.8809 |
| FLOSIM3D [77] | 0.6608 | 0.6697 | 0.7105 | 0.7234 | 0.7506 | 0.7568 | 0.7008 | 0.7131 | 0.7449 | 0.7499 | 0.6358 | 0.6432 | 0.7208 | 0.7312 |

**Table VI.** PCC values for different NR quality metrics and for each specific type of distortion

| Quality Metric | Additive White Gaussian Noise | | Simulcast Compression | | Blurring | | Brightness Shift | | 3D Video Compression | | View Synthesis | | Depth Map Compression | |
|---|---|---|---|---|---|---|---|---|---|---|---|---|---|---|
| Distortion | Original | Saliency Inspired | Original | Saliency Inspired | Original | Saliency Inspired | Original | Saliency Inspired | Original | Saliency Inspired | Original | Saliency Inspired | Original | Saliency Inspired |
| IQVG [25] | 0.5568 | 0.5609 | 0.6811 | 0.6899 | 0.6930 | 0.7003 | 0.4881 | 0.5008 | 0.6500 | 0.6626 | 0.5579 | 0.5713 | 0.6883 | 0.6940 |
| GBIM [53] | 0.5253 | 0.5745 | 0.5979 | 0.6340 | 0.6062 | 0.6439 | 0.4790 | 0.5123 | 0.6011 | 0.6341 | 0.5719 | 0.6068 | 0.6162 | 0.6549 |
| NRPBM [54] | 0.5120 | 0.5579 | 0.5977 | 0.6449 | 0.6060 | 0.6557 | 0.4945 | 0.5347 | 0.6034 | 0.6380 | 0.5774 | 0.5999 | 0.5665 | 0.5879 |
| Q_blur_Farias [26] | 0.5989 | 0.6020 | 0.6229 | 0.6299 | 0.6656 | 0.6702 | 0.5898 | 0.5911 | 0.6345 | 0.6389 | 0.6112 | 0.6178 | 0.6157 | 0.6201 |
| Q_block_Farias [26] | 0.5789 | 0.5888 | 0.6459 | 0.6512 | 0.6450 | 0.6498 | 0.5678 | 0.5703 | 0.6335 | 0.6377 | 0.6047 | 0.6127 | 0.6569 | 0.6728 |
| Q_Sadaka [27] | 0.6127 | 0.6187 | 0.6984 | 0.7035 | 0.7022 | 0.7093 | 0.6356 | 0.6397 | 0.6625 | 0.6691 | 0.6339 | 0.6514 | 0.6674 | 0.6722 |
| VQSM [28] | 0.6454 | 0.6556 | 0.6871 | 0.6909 | 0.7000 | 0.7093 | 0.6339 | 0.6421 | 0.6798 | 0.6874 | 0.6567 | 0.6655 | 0.6874 | 0.6910 |
| AQI [55] | 0.6623 | 0.6931 | 0.6712 | 0.7129 | 0.6822 | 0.7257 | 0.6345 | 0.6568 | 0.6683 | 0.6998 | 0.6110 | 0.6456 | 0.6892 | 0.7212 |
| QA3D [52] | 0.6892 | 0.7125 | 0.7265 | 0.7676 | 0.7592 | 0.7933 | 0.6887 | 0.7020 | 0.7121 | 0.7449 | 0.6886 | 0.7339 | 0.6679 | 0.7009 |
| NOSPDM [29] | 0.7467 | 0.7563 | 0.8089 | 0.8165 | 0.8157 | 0.8207 | 0.7369 | 0.7440 | 0.7773 | 0.7811 | 0.7507 | 0.7579 | 0.7834 | 0.7930 |
| Q_Ryu [30] | 0.7611 | 0.7698 | 0.8671 | 0.8747 | 0.8612 | 0.8808 | 0.6912 | 0.7023 | 0.8081 | 0.8156 | 0.6729 | 0.6804 | 0.7511 | 0.7593 |
| APT [78] | 0.5997 | 0.6033 | 0.6378 | 0.6437 | 0.6665 | 0.6727 | 0.5987 | 0.6003 | 0.6985 | 0.7098 | 0.7074 | 0.7096 | 0.6712 | 0.6789 |

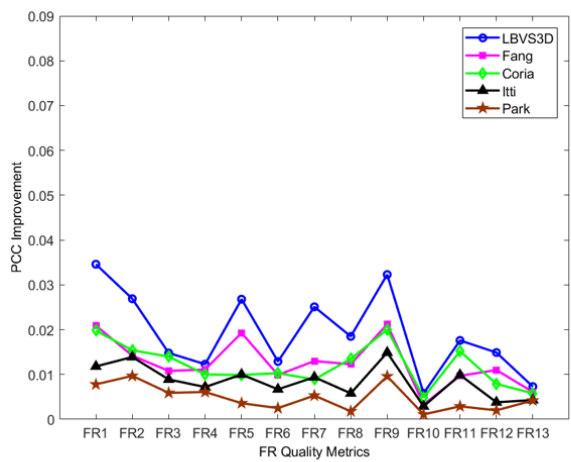
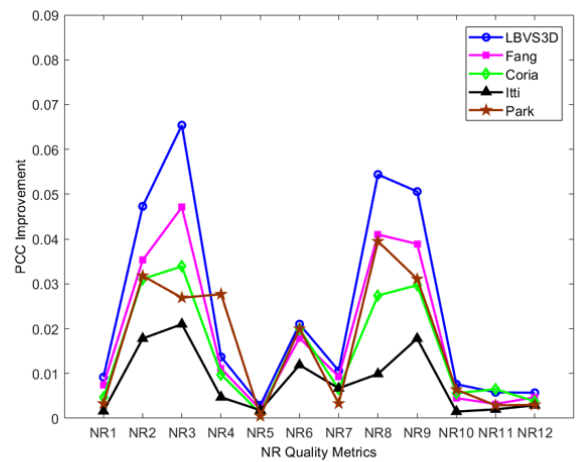

(a)    (b)

**Fig. 6.** Resulting PCC improvements when saliency maps of different VAMs are integrated to: (a) FR and (b) NR quality metrics. FR metrics from left to right: PSNR, SSIM [31], MS-SSIM [50], VIF [51], Ddl1 [42], OQ [43], CIQ [44], PHVS3D [45], PHSD [46], MJ3D [47], Q_Shao [48], HV3D [49], FLOSIM3d [77], and NR metrics from left to right: IQVG [25], GBIM [53], NRPBM [54], Q_blur_Farias [26], Q_block_Farias [26], Q_Sadaka [27], VQSM [28], AQI [55], QA3D [52], NOSPDM [29], Q_Ryu [30], APT [78].

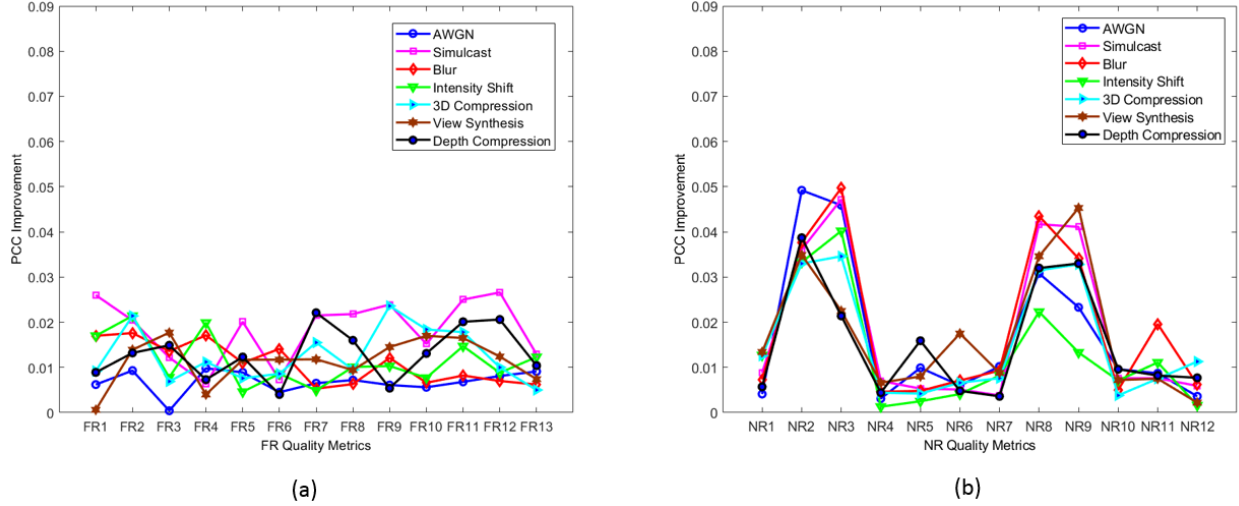

**Fig. 7.** Resulting PCC improvements when saliency maps of LBVS3D [38] VAM are integrated to: (a) FR and (b) NR quality metrics. FR metrics from left to right: PSNR, SSIM [31], MS-SSIM [50], VIF [51], Ddl1 [42], OQ [43], CIQ [44], PHVS3D [45], PHSD [46], MJ3D [47], Q_Shao [48], HV3D [49], FLOSIM3D [77], and NR metrics from left to right: IQVG [25], GBIM [53], NRPBM [54], Q_blur_Farias [26], Q_block_Farias [26], Q_Sadaka [27], VQSM [28], AQI [55], QA3D [52], NOSPDM [29], Q_Ryu [30], APT [78].

**Table VII.** Average PCC improvement for FR and NR quality metrics and for each specific type of distortion

| Quality Metric | Distortion | Additive White Gaussian Noise | Simulcast Compression | Blurring | Brightness Shift | 3D Video Compression | View Synthesis | Depth Map Compression |
|---|---|---|---|---|---|---|---|---|
| **FR** | | 0.0067 | 0.0184 | 0.0109 | 0.0114 | 0.0127 | 0.0111 | 0.0128 |
| **NR** | | 0.0170 | 0.0181 | 0.0190 | 0.0132 | 0.0158 | 0.0174 | 0.0154 |

Other observations from Fig. 6:
- For both FR and NR metrics, VAMs from Park et. al [63] and Itti et. al [64] have shown to result in less *PCC* improvements. This might be due to the fact that they are less accurate in the presence of distortions, and in general have lower saliency prediction performance for 3D video [41].
- Repeating patterns in the curves from different VAMs in Fig. 6 suggests that different VAMs show a kind of agreement in providing additional *PCC* performance improvements when integrated to different FR and NR quality metrics. The improvement amplitude however varies from one VAM to another.
- In the case of FR metrics, *PSNR*, *Ddl1*, *CIQ*, *PHSD* have received highest *PCC* improvements when saliency predictions are integrated to them.
- In the case of NR metrics, *GBIM, NRPBM, AQI*, and *QA3D* have received the highest *PCC* improvements.

We further study the performance improvements by using stereo saliency information for each type of distortion separately. Table V, Table VI, and Table VII contain the *PCC* values for FR and NR metrics before and after incorporation of stereo saliency maps. Fig. 7 shows the *PCC* improvements in the case of FR and NR quality assessment for different kinds of distortions. It is observed from Table V, Table VI, Table VII, and Fig. 7 that different kinds of distortions receive a roughly similar amount of improvements, except AWGN for FR case, which could be due to less accurate disparity map estimation from the distorted videos for this type of distortion.

Other observations from Table V, VI, VII, and Fig. 7:
- NR metrics show a higher degree of consistency in improvements they receive from saliency information over different kinds of distortions than FR metrics. In other words, regardless of the distortion type, NR metrics seem to show similar amount of improvement in *PCC*.
- For NR metrics, 'intensity shift' receives the least overall *PCC* improvements, perhaps due to the fact that the NR metrics are not designed to measure this kind of distortion.

In addition to the observations regarding the saliency integration improvements, we can deduct from Table III that *HV3D* [49], *MJ3D* [47], and *Q_Shao* [48], deliver superior FR 3D video quality assessment performance in comparison to the other FR metrics. In the case of NR 3D metrics, *Q_Ryu* [30], *NOSPDM* [29], and *QA3D* [52] demonstrate superior quality assessment performance. Moreover, regarding the strengths and weaknesses of individual metrics, the following facts are drawn from Table V and Table VI:

- *PSNR* is particularly bad at measuring blur in 3D, whereas other FR metrics are really good for this particular distortion.
- Most of the FR metrics perform well for 3D video compression quality measurement.
- Except *PSNR* and *SSIM*, other FR metrics show a fair performance on depth map compression artifacts. This is due to the fact that these 2D metrics do not monitor the depth and 3D related aspects of video.
- Although a 2D FR metric, *VIF* demonstrates a fair *PCC* performance for various kinds of distortions. This is because 3D quality is largely defined by 2D quality and *VIF* performs well for 2D stimulus.
- Interestingly, some NR metrics have a better performance in 3D quality assessment than some of the FR ones. This is due to the fact that NR metrics are generally designed having a specific target application in mind (e.g. assessing blockiness or blur), and generally successful in doing so.
- Both FR and NR metrics seem to perform poorly in general for intensity shift artifact, most likely because it's not a common distortion so not many metrics are designed to handle it.
- NR metrics show less consistency for handling different kinds of distortions. This is expected due to the fact that NR metrics are usually targeted to a specific type of distortion.

## V. CONCLUSION

In this paper we study the added value of using stereo saliency prediction in full-reference and no-reference quality assessment tasks. To this end, we leverage the stereo saliency prediction results to modify FR and NR quality metrics and re-evaluate their performance. We measure the performance improvements using a large database of stereoscopic videos with several representative types of distortions. Performance evaluations revealed that using stereo saliency in general improves the quality assessment accuracy. However, the improvements are more significant in the case of NR video quality assessment.

## VI. FUTURE WORKS

Future works include investigating the possibility of integrating emotional features in the overall 3D video quality of experience. It is known that images/videos can affect people on an emotional level [67]-[68]. Since the emotions that arise in the viewer of an image can highly impact the viewer's quality of experience, it is necessary to investigate if quality metrics can be integrated with emotion detection features (as well as VAMs). In the case of 3D video this can be different than the 2D case.